\documentstyle[12pt,lnfprep,epsfig,graphicx]{article}

\def\FI{$\phi \,$}
\def\lambda{$\Lambda \,$}

\newcommand{\Header}{
  \begin{tabular}{rl}
  \hspace{-.4cm}\includegraphics{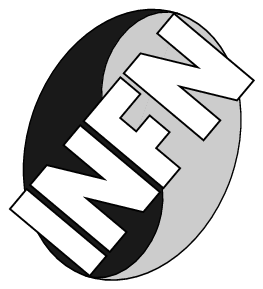} &
    \renewcommand{\arraystretch}{0.5}
    \begin{tabular}{r}
      {\hspace{1cm}~\LARGE\sffamily LABORATORI~ NAZIONALI~ DI~ FRASCATI}\\
      \\
      {\Large\sffamily SIS-Pubblicazioni}\\
    \end{tabular}
    \renewcommand{\arraystretch}{1}
  \end{tabular}
  \vskip 1cm
  \begin{flushright}
  \renewcommand{\arraystretch}{0.5}
    \begin{tabular}{r}
      {\underline{LNF-01/026(P)}}\\    
      {\small 8 Novembre 2001} \\      
      \\
    \end{tabular}
  \end{flushright}
  \renewcommand{\arraystretch}{1}
  \vskip 1 cm
  }
\begin{document}
\begin{titlepage}
\title{ 
  \Header
  {\large \bf THE FINUDA EXPERIMENT: STATUS AND PERSPECTIVES}
}
\author{
M. Bertani  \\
\it{INFN, Laboratori Nazionali di Frascati, P.O. Box 13,
I-00044 Frascati, Italy}\\
on behalf of the FINUDA Collaboration
} 
\maketitle
\baselineskip=14pt

\begin{abstract}  
FINUDA is a hypernuclear physics experiment that will be carried out 
at DA$\Phi$NE, 
the $e^+e^-$  \FI-factory currently in operation at the INFN Frascati
Laboratory.
The apparatus,  which is assembled in the DA$\Phi$NE hall, 
consists of a magnetic spectrometer with high resolution 
tracking capabilities. In this paper
the status of the experiment is presented, together with
the main features of the apparatus and of its physics program.
\end{abstract}


\vspace*{\stretch{2}}
\begin{flushleft}
\end{flushleft}
\begin{center}
{\it Presented at the 9th International Symposium
on Meson-Nucleon Physics and the Structure of the Nucleon\\
26-31 July, 2001  Washington, DC, USA} \\
\end{center}
\end{titlepage}
\pagestyle{plain}
\setcounter{page}2
\baselineskip=17pt
\section{Introduction}
An hypernucleus is a many-body system composed of conventional (non-strange)
nucleons and one or more hyperons ($\Lambda, \, \Sigma$ or $\Xi$). 
The presence of 
the strangeness degree of freedom in a hypernucleus adds a new 
dimension to the evolving picture of nuclear physics.
\par
The FINUDA Collaboration~\cite{finuda} has a very ambitious program in 
hypernuclear
physics including high statistics and high resolution spectroscopy,
study of hypernuclear mesonic and non-mesonic decay modes,
search for $\Sigma$-hypernuclei and neutron-rich
hypernuclear states.
\par
The peculiar idea of the FINUDA experiment is to stop the 
large flux of slow and monochromatic $K^-$ (127~MeV/c)
coming from the main  \FI decay $ \Phi \rightarrow K^+ K^- \, \, (49\%)$¸
in thin nuclear targets ($0.1 \div 0.3 \, \rm{g \, cm^{-2}}$) 
with minimum straggling.
After degradation and nuclear capture \lambda-hypernuclei are 
produced through the reaction:
\begin{equation}
K^-_{stop} + {~^A{Z}} \rightarrow \, {~^{A}_{\Lambda}{Z}} + \pi^-
\label{hyp}
\end{equation}
and the spectroscopy of hypernuclear states can be performed by 
measuring the
momentum of the isotropically emitted $\pi^-$. This feature provides 
unprecedented momentum resolution, as long as transparent detectors are
employed before and after the target.
\par
In the case of $\Lambda$ hypernucleus formation, the
following weak-interaction non mesonic decay modes of the $\Lambda$
are strongly favored in medium-heavy nuclei:
\begin{equation}
\Lambda + n \rightarrow n + n \quad \quad \quad , \quad  \quad \quad 
\Lambda + p \rightarrow n + p
\label{decay}
\end{equation}
which are interesting for studying the validity of the empirical
$\Delta I=1/2$ rule.
\par  
The FINUDA magnetic spectrometer has the typical cylindrical geometry of 
collider experiments, therefore it is capable of detecting the $\pi^-$
from hypernuclear  formation, eq.~\ref{hyp}, in coincidence  with the products of 
the \lambda decay, eq.~\ref{decay}. 
Up to now, this is a unique capability in the hypernuclear physics panorama.
%
%
\section{The FINUDA spectrometer}
FINUDA is a high resolution spectrometer 
($\Delta p/p \simeq 0.3\% \,\rm{FWHM}$)with cylindrical geometry,
characterized by a large solid-angle
($\sim 70\%$ of $4\pi$), good triggering capabilities, 
state-of-the-art tracking,
particle identification and neutron detection. 
\par
\begin{figure}[t]
\centerline{\epsfig{file= 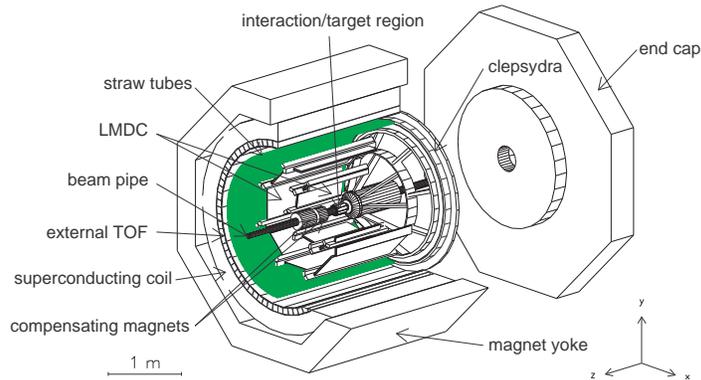,width=.65\textwidth,silent=,clip=}}
\caption{
Layout of the FINUDA apparatus.}
\label{det}
\end{figure}

  
The apparatus, described in detail in~\cite{finuda,giano} and references 
therein, is sketched in fig.~\ref{det} and consists 
of an inner section surrounding the
interaction region (beam pipe,  thin
scintillator counter barrel, 8-fold nuclear target, silicon microstrip
detectors), an external tracker
(low-mass planar drift chambers (LMDC) and a straw tube array detector),
an outer scintillator array and a superconducting solenoid 
providing a maximum
magnetic field of 1.1\,T. The whole tracking
volume ($8~\rm{m^3}$) is
immersed in a He atmosphere to minimize Multiple Coulombian Scattering.
The geometry of the spectrometer, the position of the detectors and the value 
of the maximum magnetic field have been optimized for maximizing 
the momentum resolution and acceptance for the prompt $\pi^-$ from 
hypernuclear formation (eq.~\ref{hyp}). For such  $\pi^-$ (250-280 MeV/c), a 
momentum resolution of 0.3\%(FWHM) is obtained, corresponding to a 
resolution of 700 KeV in the hypernuclear energy levels.
\par
%
\section{The physics program: expected performances}
FINUDA will investigate a wide physics program 
consisting both of hypernuclear spectroscopy and of hypernuclear 
decays~\cite{finuda,bert,felic,patic}. 
High statistics p-shell hypernuclear studies are foreseen: 
$^{12}_{\Lambda}C$,
$^7_{\Lambda}Li$, $^9_{\Lambda}Be$, $^{10}_{\Lambda}B$ and the light 
hypernuclei $^{5}_{\Lambda}He$, $^{4}_{\Lambda}He$ and  $^{4}_{\Lambda}H$  
will be produced using a $^6Li$ target.
\par
\begin{figure}[t]
\begin{centering}
\includegraphics[width=8.7cm,angle=90]{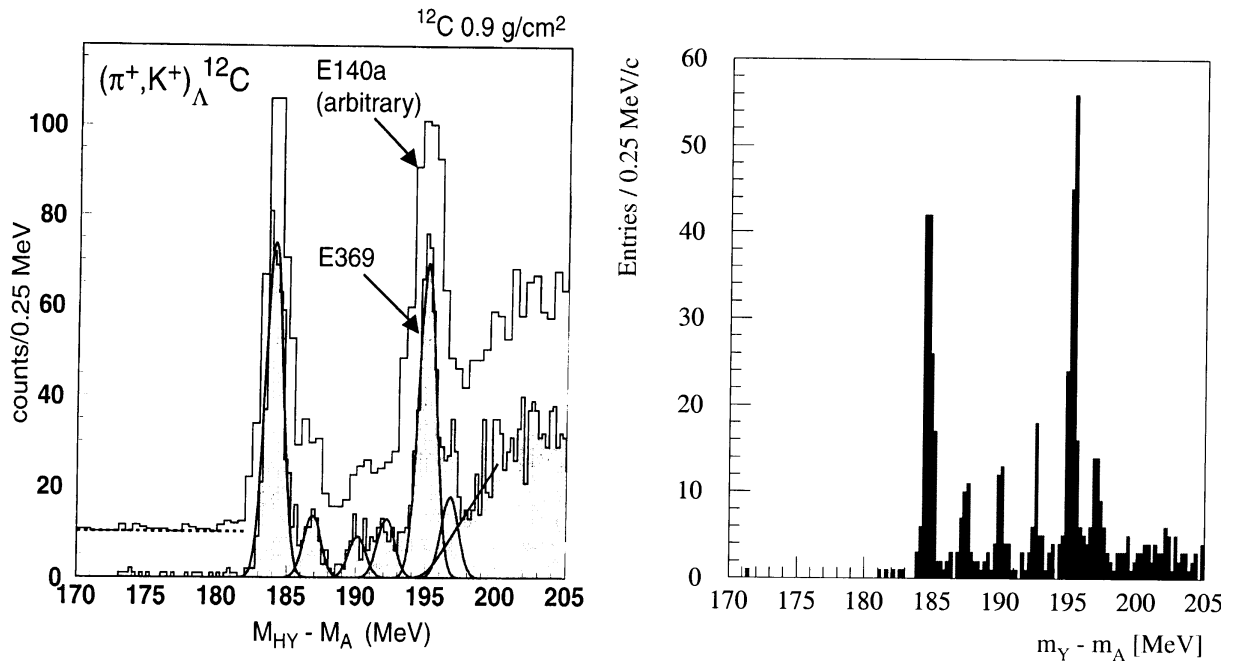}
\caption{
Left (from ref.~\cite{e369}): hypernuclear mass spectrum of $^{12}_{\Lambda}C$
from KEK experiments E369 and previous E140a with energy resolution 
of 1.45~MeV and 1.9~MeV (FWHM) respectively.  
Right~:~FINUDA simulation of the E369 $^{12}_{\Lambda}C$
levels reconstructed  with an energy resolution of 700~KeV (FWHM).}
\label{kek-fin}
\end{centering}
\end{figure}

Fig.~\ref{kek-fin}~(left side) shows 
the hypernuclear mass spectrum of $^{12}_{\Lambda}C$
recently measured by experiment E369~\cite{e369} at KEK with an energy resolution
of 1.45~MeV (FWHM). An old spectrum form the previous E140~\cite{e140} experiment
with an energy resolution of 1.9~MeV is superimposed in an arbitrary scale for comparison. 
This picture shows how the improvement in  resolution is essential for the
understanding of hypernuclear spectra. 
The E369 hypernuclear levels have been simulated 
with the FINUDA Monte Carlo and injected in the reconstruction program
to test the physics performances
of the apparatus.   
The result is a very clean spectrum with practically no background
(right side of fig.~\ref{kek-fin}): FINUDA with 
700~KeV (FWHM) energy resolution may reveal finer splittings in the same 
hypernuclear spectra. The simulated spectrum corresponds to an integrated 
luminosity of $5 \, \rm{pb^-1}$, that is about 2 days at the present DA$\Phi$NE 
luminosity, $3 \times 10^{31} {\rm cm^{-2} s^{-1}}$.
\par
Concerning the weak decay studies of hypernuclei, FINUDA has the 
opportunity of measuring 
with good accuracy the ratio $\Gamma_n/\Gamma_p$  between the $p$-induced and
$n$-induced hypernuclear weak decay amplitudes (eq.~\ref{decay}) which is
related to the validity of the $\Delta I=1/2$ rule.  In fact there are indications~\cite{deltai} that 
the rule may be violated in the non mesonic decay modes. As it has
already been noted, the hypernuclear formation and decay products 
can be measured in coincidence and within the same apparatus. 
This double capability of FINUDA spectrometer 
is up to now unique with respect to existing hypernuclear facilities.
\par  
Table~\ref{tab} shows
the expected FINUDA performances in hypernuclear high resolution  
\-spectro\-scopy and weak decay studies for the $^{12}_{\Lambda}C$ which 
is the best known hypernuclear system
and can be used as a reference mark. 
The table refers an integrated luminosity of 
$50\,\rm{pb^{-1}}$, corresponding to about 20 days of data taking at 
the present DA$\Phi$NE luminosity and to a formation rate of the 
$^{12}_{\Lambda}C$ ground state of $10^{-3}$ per stopped $K^-$.
The comparison with existing measurement (\-se\-cond column of tab.~\ref{tab})
shows that, even with a reduced
luminosity with respect to DA$\Phi$NE and  FINUDA projects, FINUDA can in
principle significantly lower the statistical error: in some cases the 
reduction is even of one order of magnitude. In particular the 
accuracy that will be
reached in the $\Gamma_p / \Gamma_n$ ratio measurement will hopefully allow 
to discriminate among different theoretical predictions. 
\begin{table}[t]
\begin{center}
\begin{tabular}{|c|c|c|c|}
\hline
\hline
observable& B.R.($\%$) $^{12}_{\Lambda}C$ g.s. & collected event& stat. err. ($\%$)\\
\hline
\hline
high resolution &    &  $11.2 \times 10^3$  &         \\
hypernuclear spectroscopy    &   &                     &          \\
\hline
$\Gamma_{tot}/\Gamma_{\Lambda}$ & $1.25 \pm 0.18$\cite{szy} & 
$2.2\times 10^3$ & $\sim 2$ \\
\hline
$\Gamma_{\pi^-}/ \Gamma_{\Lambda}$ & $0.14 \pm 0.07 \pm 0.03$ \cite{Noumi} & 
$7.0\times 10^2$ & $\sim 4$ \\
\hline
$\Gamma_p/\Gamma_{\Lambda}$ &  $0.31 \pm 0.07 ^{+0.11}_{-0.04}$\cite{Noumi} & &\\
only $p$ detected & & $2.2\times 10^3$  & $\sim 2$  \\
both $p$ and $n$ detected & & $1.9\times 10^2  $  &  $\sim 7$  \\	    
\hline
$\Gamma_n/\Gamma_{\Lambda}$ &      & 96 & $\sim 10$ \\
both $n$ detected & & & \\
\hline
$\Gamma_n/\Gamma_p$ &  $1.87 \pm 0.59 ^{+0.32}_{-1.00}$\cite{Noumi} & & \\
                   & $1.33 ^{+1.12}_{-0.81}$\cite{szy} & & $\sim 10$  \\
		   & $0.59^{+0.17}_{-0.14}$\cite{mont} & & \\
\hline        
\end{tabular}
\label{tab}
\caption{Finuda expected performances in hypernuclear spectroscopy and decay 
studies for $^{12}_{\Lambda}C$, for an integrated luminosity 
of $50\,\rm{pb^{-1}}$.}
\end{center}
\end{table}
%
%
\section{Present status and future plans}
DA$\Phi$NE~\cite{dafne}, the $e^+e^-$  Frascati \FI-factory, 
has been designed to achieve high luminosity 
\linebreak{($1 \div  5 \times 10^{32} {\rm cm^{-2} s^{-1}}$)} at the
\FI resonance energy (1.020~GeV) in two opposite interaction regions
where the KLOE~\cite{kloe} and the DEAR~\cite{dear}
detectors are currently positioned. FINUDA
spectrometer will take the place of DEAR in the second interaction 
region. DA$\Phi$NE commissioning started in 1998 and is
now providing collisions to the KLOE and DEAR experiments
with a maximum luminosity of about $3 \times 10^{31} {\rm cm^{-2} s^{-1}}$.
\par
FINUDA magnet is in the DA$\Phi$NE pit, next to the beam
line,  since 1998; all the subdetectors are ready and tested since 1999, 
they have all been installed and tested again inside the 
superconducting magnet during
some machine shutdowns between the end of the year 2000 and August 2001.  
The roll-in of the apparatus in the second DA$\Phi$NE interaction region
is scheduled for August 2002, and consequently 
data taking with colliding beams should finally start.
\par
The FINUDA Collaboration plans to begin data taking using two sets of 
different nuclear targets, namely 4 $^{12}C$ and 4 $^{6}Li$ targets.
The first set will be a sort of reference mark, being the 
$^{12}_{\Lambda}C$ the best known hypernuclear system, 
the second set will allow the study of the light hypernuclei 
$^{5}_{\Lambda}He$, $^{4}_{\Lambda}He$, and  $^{4}_{\Lambda}H$~\cite{felic}.
\section{Conclusions}
A powerful spectrometer for high quality hypernuclear studies is ready to 
start taking data at the Frascati DA$\Phi$NE \FI-factory. 
It is scheduled to be rolled in the machine second interaction region in August 2002. 
As shown in table~\ref{tab}, even with a reduced luminosity with respect to DA$\Phi$NE and FINUDA designs, 
FINUDA can give world class results in hypernuclear spectroscopy and weak decay measurements thus allowing 
to discriminate among the different theoretical predictions.
%

\end{document}